# A COMPARATIVE STUDY ON REMOTE TRACKING OF PARKINSON'S DISEASE PROGRESSION USING DATA MINING METHODS

Peyman Mohammadi[1], Abdolreza Hatamlou[2] and Mohammad Masdari[3]

[1] Department of Computer Engineering, Science and Research Branch, Islamic Azad University, West Azerbaijan, Iran

[2]Islamic Azad University, Khoy Branch, Iran

[3]Department of Computer Engineering, Urmia Branch, Islamic Azad University, Urmia, Iran

## ABSTRACT

*In recent years, applications of data mining methods are become more popular in many fields of medical diagnosis and evaluations. The data mining methods are appropriate tools for discovering and extracting of available knowledge in medical databases. In this study, we divided 11 data mining algorithms into five groups which are applied to a dataset of patient's clinical variables data with Parkinson's Disease (PD) to study the disease progression. The dataset includes 22 properties of 42 people that all of our algorithms are applied to this dataset. The Decision Table with 0.9985 correlation coefficients has the best accuracy and Decision Stump with 0.7919 correlation coefficients has the lowest accuracy.*

## KEYWORDS

## 1. INTRODUCTION

The Parkinson's Disease (PD) is a chronic neurological disorder with an unknown etiology, which usually affects people over 50 years old [1]. It was reported that the Parkinson, beyond many unknown cases, has affected one million people [2]. It is expected that as the age of earth's population increases, the number of patients with PD also increases.

The number of Parkinson's disease (PD) patients is estimated to be 120–180 out of every 100,000 people, although the percentage (and hence the number of affected people) is increasing as the life expectancies increase. The causes of Parkinson's disease (PD) are unknown; however, researches have shown that the degradation of the dopaminergic neurons affects the dopamine production [3].The PD symptoms include limb tremor (especially at rest), muscles stiffness and movement slowness, difficulty in walking, balance and coordination (especially when beginning of motion), difficulty in eating and swallowing, and vocal disorder and mood disorders [4, 5, 6]. This disease causes voice disorder in 90% of patients [7]. Dysarthria is a motor speech disorder that indicates inability in expression properly.

It's observable in patients with PD to have hypokinetic dysarthria, and its classical symptoms include reduced vocal loudness (hypophonia), monopitch, disruption of voice quality, and abnormally fast rate of speech [8]. For  decades, researchers have strived to understand more about this disease and therefore to find some methods for successfully limiting its symptoms which are most commonly periodic muscle tremor and/or rigidity. Other symptoms such as akinesia, bradykinesia, and dysarthria may occur in later stages of PD [9].

In recent years, development and innovation in fusion data equipment such as sensors in scientific and medicine areas has led to production of massive data values. The modern medicine produces massive amounts of data stored in databases whose use of these large values of data is time-consuming and in some cases it is impossible. In this regard, we can use some data mining methods to perform this difficult task. Of course, all data mining methods have their own limitations and constraints and are our task to compare methods and algorithms to choose the best and most useful method.

The term used as "Data Mining", was introduced in the 1990s; however data mining is the progress and improvement of a field with a long history [10]. As datasets are growing in size, applications and complexity, Direct hands-on data analysis has increasingly been augmented with indirect, automatic data processing. This has been achieved by other discoveries in computer science such as artificial neural networks (ANNs), classification and clustering algorithms plans. Data mining techniques and algorithms has been accomplished for genetic algorithm (GA) in the 1950s, and for decision trees (DTs) in the 1960s. Also it's a supported vector machine (SVM) in 1990s methods [11]. The first use of data mining techniques in health information systems (HIS) and clinical application was fulfilled with the expert systems developed since the 1970s [12].

Given to knowledge, experience and the sense of physicians, medicine decisions and reviews are done, and because of massive amount of medicine data which are kept in medical clinics and health centers, the data mining methods can be appropriate tools for discovering available knowledge in this database to help the physicians with diagnosis, treatment and tracking process.

The rest of this paper is organized as follows. First explained data mining techniques and algorithms used in this study in the next section; and data mining application in healthcare field are described in Section 3. Then the data mining algorithms are categorized and applied to a special dataset in Section 4. The results are analysed and discussed in Section 5. Finally, some concluding remarks and proofs outcome as well as future research lines are presented in Section 6.

## 2. DATA MINING

Database pertaining to trade, agriculture, cyberspace, details of phone calls, medical data and etc. are collected and stored very rapidly with the development of information technology, data collection and production methods. So, since three decades ago, the human was thinking about a new method to access hidden data in this huge and large database because the traditional systems were not able to do so. Competition in economic, scientific, political, military fields and the importance of access to useful information among a high amount of data without human intervention caused the data analysis science or data mining to be established.

The concept of data mining has been developed in the late 1980s, which went further during 10 years later [13]. Data mining is the process of discovering meaningful new correlations, patterns and trends by sifting and investigating through large amounts of data stored in repositories, using pattern recognition technologies as well as statistical and mathematical techniques (Gartner Group). As like as mining gold through huge rocks and large amount of soils, data mining is extracting beneficial knowledge from bulky data sets. But they aren't entirely the same; there is a fine point which is data mining but it isn't just searching and gaining knowledge, it is to obtain knowledge by analysing and reconsidering [14].

Many people consider data mining as equivalent of another commonly used term, Knowledge Discovery from Data, or the KDD. Alternatively, others view data mining as simply an indispensable step in the process of knowledge discovery [12].

Typically, the steps of knowledge discovery (KD) are divided into 7 phases. Four primary steps are pre-processing data and different forms of data preparation for data mining. As mentioned above, data mining as a fifth steps, is one of the necessary steps of this process. Finally, two last steps have the task of identifying useful patterns and displaying it to user. Below there is a brief description of these steps:

- Data cleaning: clean incompatible data and noises.
- Data integration: integrate multiple sources, if any.
- Data selection: restore data related to analysis and evaluate database.
- Data transformation: transform or modulate data into searchable forms by summarizing or aggregating.
- Data mining: process of applying intelligent algorithms to exploiting data patterns.
- Pattern evaluation: identify related patterns.
- Knowledge presentation: present extracted knowledge to users using presentation techniques and visual modeling.

Therefore data mining is a step in the KDD process consisting of a particular enumeration of patterns over the data, subjected to some computational limitations. The term pattern goes beyond its traditional concept and notion to include structures or models in the data warehouses. Historical data are used to discover regularities and improve future decisions [15].

The goals of data mining are to divide it into two general types as predictive and descriptive [12]. By application of the predictive type, we understand that application of the model which attempts to 'predict' the value that a certain variable may take ,given what we know at present [16]. Descriptive data mining characterizes the generic attributes of the data in the database [12].

# 3. DATA MINING IN HEALTHCARE

Today, in medical areas, data collection about different diseases is very important. Medical centers with different purposes are to collect these data. To survey these data and to obtain useful results and patterns in relation to disease is one of the objectives of the use of these data. Collected data volume is very high and we must use data mining techniques to obtain desired patterns and results among massive volume of data.

Medical and health areas are one of the most important sections in industrial societies. The extraction of knowledge among massive volume of data related to diseases and people medical records using data mining process can be lead to identifying the laws governing the creation, development and epidemic diseases. We can allow expertise and health area staffs to access these valuable data according to the environmental factors in order to identify diseases causes, anticipate and treat the diseases. Finally, this means extending the life and comforting people in the community.

For example some medicine applications of data mining are:

- Predicting health care costs.
- Determination of disease treatment.
- Analyzing and processing medical images.

- Analyzing health information system (HIS).
- Effect of drugs on the disease and its side effects.
- Predicting the success rate of medicine operations like surgeries.
- Diagnosing and predicting of most kind of diseases such as cancer.

In a study published in 2011, Cheng-Ding Chang *et al*. used data mining techniques for predicting the incidental multi-disease (hypertension and hyperlipidemia). They presented a two-phase analysis procedure to simultaneously predict both hypertension and hyperlipidemia. Firstly they chose six data mining approaches, and then they used these algorithms to select the individual risk factors of these two diseases, afterward determined the common risk factors using the voting principle. Next, they used the Multivariate Adaptive Regression Splines (MARS) method to construct a multiple predictive model for hypertension and hyperlipidemia. This study used data from a physical examination center database in Taiwan that included 2048 subjects. The proposed multi-disease predictor method in this study has a classification accuracy rate of 93.07% [17].

Also Ömer Eskidere and two other colleagues studied the performance of Support Vector Machines (SVM), Least Square Support Vector Machines (LS-SVM), Multilayer Perceptron Neural Network (MLPNN), and General Regression Neural Network (GRNN) regression methods in application to remote tracking of Parkinson's disease progression. It is found that using LS-SVM regression with log-transformation, the estimated motor-UPDRS is within 4.87 points (out of 108) for log-FS-4 feature set and total-UPDRS is within 6.18 points (out of 176) for Log-FS-4 feature set of the clinicians' estimation [18].

In one of the last-mentioned ascertainments in current year (2013), Daniel Ansari *et al*. enquired about Artificial Neural Networks (ANNs) as nonlinear pattern recognition techniques that can be used as a tool in medical decision making. Based on ANNs, by using clinical and histopathological data from 84 patients a flexible nonlinear survival model was designed. After the use of ANNs for predicting survival in pancreatic cancer, the C-index was 0.79 for the ANN and 0.67 for Cox regression [19].

# 4. PROPOSED METHODS AND IMPLEMENTATION

Data mining has wide and common uses in classification and recognition of the problems of health systems. In this section, we describe the materials and methods which are used in current research. Depending on the situation and desired outcome, there are various types of data mining algorithms that you can use. All experiments described in this paper were performed using libraries from Weka 3.7.9 machine learning environment. Before building models, the data set were randomly split into two subsets: 75% (n=4406) of the data for training set and 25% (n=1469) of the data for test set and all of data mining methods apply to total-UPDRS property.

## 4.1. Data Set

Parkinson's telemonitoring dataset was created by Athanasios Tsanas and Max Little of the University of Oxford, in collaboration with 10 medical centers in the US and Intel Corporation which developed the AHTD telemonitoring device to record the speech signals [6]. Dataset has been made available online at UCI machine learning archive very recently in October 2009. This dataset consists of around 200 recordings per patient from 42 people (28 men and 14 women) with early-stage PD which makes a total of 5875 voice recordings. Each patient is recorded with phonations of the sustained vowel /a/. The dataset contains 22 attributes including subject number, subject age, subject gender, time interval from baseline recruitment data, motor-UPDRS, total-UPDRS, and 16 biomedical voice measures (vocal features).The vocal features in the dataset are diverse, some of them based on traditional measures (Jitter, shimmer, HNR and NHR), and some (RPDE, DFA, PPE) are based on nonlinear dynamical systems theory. In the dataset, Motor-UPDRS and total-UPDRS were assessed at baseline (onset of trial), three-and-six-month trial periods, but the voice recordings were obtained at weekly intervals. Hence both Motor-UPDRS and total-UPDRS linearly interpolated [6]. Table.1 represents baseline, after three and six months, UPDRS scores with the feature labels and short explanations for the measurement along with some basic statistics of the dataset.

Table 1.  Description of the features and UPDRS scores of the Parkinson's telemonitoring dataset.

| Description' | Feature label | Min | Max | Mean | StdDev |
|---|---|---|---|---|---|
| Clinician's motor UPDRS score, linearly interpolated | Motor-UPDRS (baseline) | 6 | 36 | 19.42 | 8.12 |
| | Motor-UPDRS (after three months) | 6 | 38 | 21.69 | 9.18 |
| | Motor-UPDRS (after six months) | 5 | 41 | 29.57 | 9.17 |
| Clinician's total UPDRS score, linearly interpolated | Total-UPDRS (baseline) | 8 | 54 | 26.39 | 11.8 |
| | Total-UPDRS (after three months) | 7 | 55 | 29.36 | 11.82 |
| | Total-UPDRS (after six months) | 7 | 54 | 29.57 | 11.92 |
| Several measures of variation in fundamental frequency | MDVP: Jitter (%) | 8E-4 | 0.1 | 0.006 | 0.006 |
| | MDVP:Jitter (Abs) | 2E-6 | 4E-4 | 4E-5 | 3E-5 |
| | MDVP:Jitter:RAP | 3E-4 | 0.057 | 0.003 | 0.003 |
| | MDVP:Jitter:PPQ5 | 4E-4 | 0.069 | 0.003 | 0.004 |
| | Jitter:DDP | 10E-4 | 0.173 | 0.009 | 0.009 |
| Several measures of variation in amplitude | MDVP:Shimmer | 0.003 | 0.269 | 0.034 | 0.026 |
| | MDVP:Shimmer (dB) | 0.026 | 2.107 | 0.311 | 0.230 |
| | Shimmer:APQ3 | 0.002 | 0.163 | 0.017 | 0.013 |
| | Shimmer:APQ5 | 0.002 | 0.167 | 0.020 | 0.017 |
| | Shimmer:APQ11 | 0.003 | 0.276 | 0.028 | 0.020 |
| | Shimmer:DDA | 0.005 | 0.488 | 0.052 | 0.040 |
| Two measures of ratio of noise to tonal components in the voice | NHR | 3E-4 | 0.749 | 0.032 | 0.060 |
| | HNR | 1.659 | 37.875 | 21.679 | 4.291 |
| A nonlinear dynamical complexity measure | RPDE | 0.151 | 0.966 | 0.541 | 0.101 |
| Signal fractal scaling exponent | DFA | 0.514 | 0.866 | 0.653 | 0.071 |
| A nonlinear measure of fundamental frequency variation | PPE | 0.022 | 0.732 | 0.220 | 0.092 |

## 4.2. Functions

### 4.2.1. Simple Linear Regression (SLR)

The simple linear regression model y = α + βx + e, with measurement error (errors-in-variables model) has been subject to extensive research in the statistical literature for over a century. It is well known that the simple linear regression model with error in the predictor x is not identifiable without any extra information in the normal case. In statistics, simple linear regression is the least squares estimator of a linear regression model with a single explanatory variable. In other words, simple linear regression fits a straight line through the set of n points in such a way that makes the sum of squared residuals of the model (that is, vertical distances between the points of the data set and the fitted line) as small as possible [20].

So in the simple linear regression model, the observations $y_i$ are generated according to:

$$y_i = \alpha + \beta x_i + e_i \quad , i=1, …, n.$$

where α is the intercept parameter, β the slope parameter, $x_i$ the explanatory variables and finally $e_i$ the error terms. If α=0 we obtain the simple linear regression model without intercept.

$$y_i = \beta x_i + e_i \qquad , i=1, \dots, n.$$

You can see a simple linear regression model in Figure.1 that shows this model's parameters.

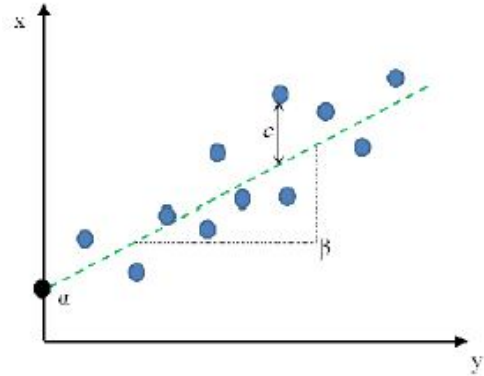

Figure 1. Simple Linear Regression Model.

### 4.2.2. Multi-Layer Perceptron (MLP)

An MLP is based on single perceptron that was introduced by Rosenblatt [21].The network contains an input layer, a hidden layer and an output layer. Each neuron at one layer receives a weighted sum from all neurons in the previous layer and provides an input to all neurons of the later layer [22].

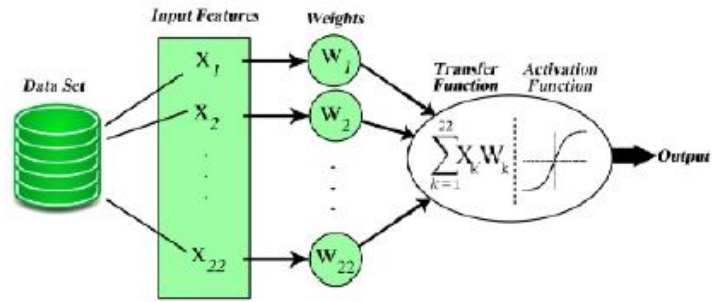

Figure 2. MLP structure that used.

However there is merely a single neuron in the output layer. The architecture of the Neural Network, used in this research, is the multilayer perceptron network architecture with 22input nodes, 13 hidden nodes in our approach. We showed the MLP structure used in this study in.

The number of input nodes is determined by the ultimate data; the number of hidden nodes is determined through trial and error; and the number of output nodes is represented as a rage, demonstrating the disease classification. Each neuron at one layer receives a weighted sum from all neurons in the previous layer and provides an input to all neurons of the later layer.

### 4.2.3. SMOreg

The Sequential Minimal Optimization (SMO) algorithm on classification tasks defined on sparse data sets has been shown to be an effective method for training SVM. SMOreg implements a sequential minimal optimization algorithm for training a support vector regression model. This implementation globally replaces all missing values and transforms nominal attributes into binary ones [23]. It also normalizes all attributes by default. We '*normalized*' training dataset used the '*polynomial*' kernel: $K(x, y) = <x, y>^p$ or $K(x, y) = (<x, y>+1)$ ^p for kernel and used '*RegSMO*' to optimizing.

## 4.3. Rules

### 4.3.1. M5Rules

Generate a decision list for regression problems used to divide and to conquer. It builds a model tree using M5 and makes the "best" leaf into a rule in each iteration of progress. The method for generating rules from model trees, called M5-Rules, is straightforward and works as follows; a tree learner is applied to the full training dataset and a pruned tree is learned. Then, the best branch is made into a rule and the tree is discarded.

All instances covered by the rule are removed from the dataset. The process is applied recursively to the remaining instances and terminates when all instances are covered by one or more rules. This is a basic divide-and-conquer strategy for learning rules; however, instead of building a single rule, as it is usually done, we built a full model tree at each stage, and make its "*best*" branch into a rule. In contrast to PART (partial decision trees), which employs the same strategy for categorical prediction, M5-Rules builds full trees instead of partially explored trees. Building partial trees leads to a greater computational efficiency, and does not affect the size and accuracy of the resulting rules [24, 25, 26, 27 and 28]. In simulating this mode we set the minimum number of instances to allow at a leaf node to 4.0.

### 4.3.2. Decision Table

The Decision Table (DT) is a symbolic way to express the test experts or design experts knowledge in a compact form [29].A decision table includes a hierarchical table. These tables consider that each entry in a higher level table gets broken down by the values of a pair of additional attributes to form another table. The structure is similar to dimensional stacking [30]. For DT implementation we used "*Best First*" method to find good attribute combinations for the decision table. For example in Table.2, the final row tells us to classify the applicant as Heart Failure (HF) if gender is Female, age is more than 42 and blood pressure isn't important.

Table 2. Classification model by decision table.

| Gender | Age | Blood of pressure | HF = Yes | HF = No |
|--------|-----|-------------------|----------|---------|
| Male | <45 | Low | - | × |
| | ≥ 45 | High | × | - |
| Female | <42 | Low | - | × |
| | ≥ 42 | - | × | - |

## 4.4. Trees

### 4.4.1. M5P

This is an algorithm for generating M5 model trees [31]. M5 builds a tree for a given instance, to predict numeric values. When the input attributes can be either discrete or continuous the algorithm requires the output attribute to be numeric. For a given instance the tree is traversed from top to bottom until a leaf node is reached. At each node of the tree a decision is made to follow a particular branch based on a test condition on the attribute associated with that node. Each leaf node has a linear regression model associated with the form:

$$w_0 + w_1 a_1 + ... + w_k a_k$$

Based on some of the input attributes $a_1, a_2, ..., a_k$ in the instance and whose respective weights $w_0, w_1, ..., w_k$ are calculated using standard regression. The tree is called a model tree as the leaf nodes contain a linear regression model to generate the predicted output [32]. We start with a set of training instances to build a model tree, using the M5 algorithm. The tree is built using a divide-and-conquer method. The M5P Model Tree algorithm in WEKA is available in the Java class ''*weka.classifiers.trees.M5P*''. The minimum number of instances to allow at a leaf node was 2.0 and after simulation number of rules was 170.

### 4.4.2. REPTree

This algorithm is a fast decision tree learner. It is also based on C4.5 algorithm and can cause classification (discrete outcome) or regression trees (continuous outcome). It uses information gain/variance to building a regression/decision tree using and prunes it using reduced-error pruning (with back-fitting). Only sorts values for numeric attributes once. Missing values are dividing by splitting the corresponding instances into pieces [33].

In implement of REPTree we set the maximum tree depth to -1, it means there was no restriction. The minimum proportion of the variance on all the data that needs to be presented at a node was 0.001, which is used for splitting, to be performed in regression trees. After implementation we have a tree whose size is 675.

### 4.4.3. Decision Stump

The decision stump is the simplest special case of a decision tree. Each decision stump consists of a single decision node and two prediction leaves. Each decision node is a rule that checks the presence or absence of specified n-gram of command sequence. The boosted decision stumps are created using a weighted voting of the decision stumps [34]. It's usually used in conjunction with a boosting algorithm, does regression or classification. Missing is treated as a separate value.

## 4.5. Lazy

### 4.5.1. IBk

This is the K-nearest neighbour classifier. We used IB1 which is the simplest instance-based learning algorithm. It gives a test instance, and to find the training instance closest to the given test instance uses a simple distance measure, and predicts the same class as this training instance. If more than one instance is the same smallest distance (multiple) to the test instance, the first one found is used.

The similarity function used here is:

$$Similarity(x, y) = -\sqrt{\sum_{i=1}^{n} f(x_i, y_i)}$$

Where the instances are described by n attributes and it's 22 in our case. We define $f(x_i, y_i) = (x_i - y_i)^2$ for numeric-valued attributes and $f(x_i, y_i) = (x_i \neq y_i)$ for Boolean and symbolic-valued attributes. The difference of IBI and nearest neighbour algorithm is that IBI normalizes its attributes ranges, processes instances incrementally, and unlike nearest neighbour algorithm has a simple policy for tolerating missing values [35]. In this model we used "*Linear NN Search*" algorithm for searching nearest neighbour which is implementing the brute force to search algorithm for nearest neighbour search in 1 nearest neighbour(s) for classification.

### 4.5.1. LWL

Lazy learning methods defer training data process until a query needs to be answered. This algorithm to assign instance weights uses an instance-based. A good choice for classification is Naive Bayes. This implementation surveys a form of lazy learning, locally weighted learning, that uses locally weighted training to average, interpolate between, extrapolate from, or otherwise combine training data [36]. In implementation of LWL model we used all of the neighbours in weighting function which is linear function and "*Linear NN Search*" algorithm for searching nearest neighbour which is implementing the brute force to search algorithm with Euclidean distance (or similarity) function.

## 4.6. Meta

### 4.6.1. Regression by Discretization

A class for a regression scheme that utilizes any distribution classifier on a version of the data that has the class attribute discretized. The predicted value based on the predicted probabilities for each interval is the expected value of the mean class value for each discretized interval. This class now also supports conditional density estimation by building a univariate density estimator from the target values in the training data, weighted by the class probabilities. To implement this model, J48 with C4.5 algorithm was adopted due to it abilities to apply negotiation strategies; and set Number of bins for discretization to 10 and estimator type for density estimating was histogram estimator. Our created J48 pruned tree size is 315 and it has 158 leaves node.

## 5. RESULTS AND DISCUSSION

We report the results from eleven methods used to study the usefulness of the data mining algorithms for remote tracking of Parkinson's disease progression. The data set, used in the detection of PD progression with eleven data mining methods, was divided into two subsets as Training and Test data. There were a total of 4406 pieces of data, exclusively used for training and 1469 pieces of data exclusively used for testing. The classification accuracies obtained by this study for PD dataset are presented in Table.3 You can compare and assimilate results of the algorithms in correlation coefficient, Mean absolute error, Root mean squared error, relative absolute error (in percentage) and root relative squared error (in percentage). In statistics, the Mean Absolute Error (MAE) is a quantity used to measure how close forecasts or predictions are to the eventual outcomes and the Root Mean Square Error (RMSE) is a frequently used measure of the differences between values predicted by a model or an estimator and the values actually observed.

Table 3. Experiential results of classification models.

| Category | Classifier | Correlation coefficient | Mean absolute error | Root mean squared error | Relative absolute error (%) | Root relative squared error (%) |
|---|---|---|---|---|---|---|
| Functions | Simple Linear Regression (SLR) | 0.9453 | 2.5483 | 3.4291 | 30.1693 | 32.6628 |
| | Multi-Layer Perceptron (MLP) | 0.9922 | 1.0866 | 1.5274 | 12.8641 | 14.5488 |
| | SMOreg | 0.9496 | 2.3315 | 3.3087 | 27.6028 | 31.5161 |
| Rules | M5Rules | 0.9967 | 0.5570 | 0.8492 | 6.5947 | 8.0889 |
| | Decision Table | 0.9985 | 0.3150 | 0.5688 | 3.7293 | 5.4183 |
| Trees | M5P | 0.9964 | 0.5741 | 0.8968 | 6.7966 | 8.5426 |
| | REPTree | 0.9969 | 0.4159 | 0.8280 | 4.9237 | 7.8867 |
| | Decision Stump | 0.7919 | 5.3167 | 6.4190 | 62.9445 | 61.1426 |
| Lazy | IBk | 0.9974 | 0.4221 | 0.7509 | 4.9973 | 7.1521 |
| | LWL | 0.8060 | 5.1674 | 6.2296 | 61.1765 | 59.3379 |
| Meta | Regression By Discretization | 0.9900 | 1.2019 | 1.4845 | 14.2289 | 14.1405 |

Figure.3 summarized correlation coefficient of each algorithm in categories. We notice Decision Table gave higher accuracy of correlation coefficient on our dataset (99.85%) in Rules category and M5Rules algorithm has comparatively a good accuracy equivalent to 99.67%.

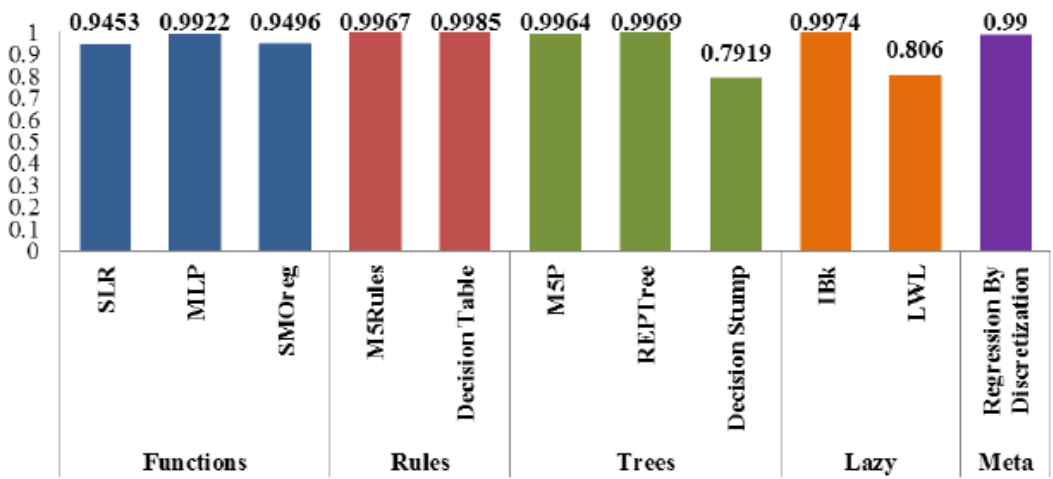

Figure 3. Correlation coefficient of classification models.

The relative absolute error is very similar to the relative squared error in the sense that it is also relative to a simple predictor, which is just the average of the actual values. In this case, though, the error is just the total absolute error instead of the total squared error. Thus, the relative absolute error takes the total absolute error and normalizes it by dividing by the total absolute error of the simple predictor and the root relative squared error is relative to what it would have been if a simple predictor had been used. More specifically, this simple predictor is just the average of the actual values. Thus, the relative squared error takes the total squared error and normalizes it by dividing by the total squared error of the simple predictor. By taking the square root of the relative squared error one reduces the error to the same dimensions as the quantity being predicted. In Figure.4 we show these two parameters in a chart.

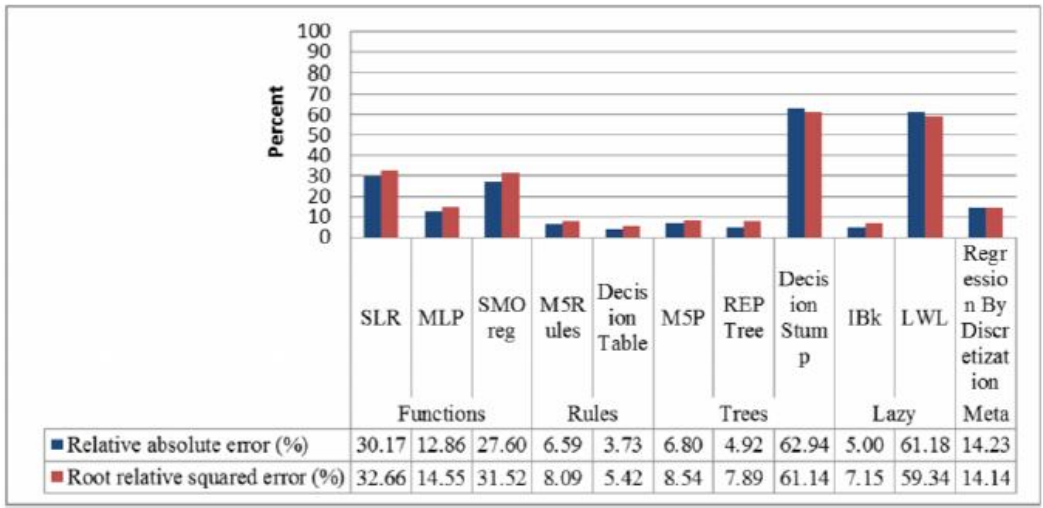

| | SLR | MLP | SMO reg | M5R ules | Decis ion Table | M5P | REP Tree | Decis ion Stum p | IBk | LWL | Regr essio n By Discr etizat ion |
|---|---|---|---|---|---|---|---|---|---|---|---|
| | Functions | | | Rules | | Trees | | | Lazy | | Meta |
| Relative absolute error (%) | 30.17 | 12.86 | 27.60 | 6.59 | 3.73 | 6.80 | 4.92 | 62.94 | 5.00 | 61.18 | 14.23 |
| Root relative squared error (%) | 32.66 | 14.55 | 31.52 | 8.09 | 5.42 | 8.54 | 7.89 | 61.14 | 7.15 | 59.34 | 14.14 |

Figure 4. Relative absolute error and Root relative squared error of classification models.

## 5. CONCLUSION AND FUTURE WORKS

This experimental study compares classification performance of different eleven data mining algorithms via using of a Parkinson's Telemonitoring dataset. This dataset comprises of 22 attributes with various ranges of values. Early detection of any kind of disease is an important factor and for PD, remote tracking of UPDRS using voice measurements would facilitate clinical monitoring of elderly people and increase the chances of its early diagnosis.

Used data mining techniques are Simple Linear Regression (SLR), Multi-Layer Perceptron (MLP), SMOreg, M5Rules, Decision Table, M5P, REPTree, Decision Stump, IBk, LWL and Regression by Discretization models. The best approach for remote Parkinson's Telemonitoring is Decision Table model also M5Rules in Rules category has a good result with correlation coefficient of 0.9967. Mathematical equations from different data mining techniques for calculating experimental results were derived.

## Authors

**Peyman Mohammadi** is an M.Sc. student in Computer Engineering Department, Science and Research Branch, Islamic Azad University, West Azerbaijan, Iran. His interested research areas are in the Artificial Neural Networks, Data Mining and Machine Learning techniques. 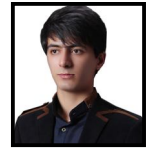